\documentclass[reprint,
superscriptaddress,
 amsmath,amssymb,
 aps,
onecolumn
]{revtex4-2}

\usepackage{cases}
\usepackage{amssymb}
\usepackage{epstopdf}
\usepackage{physics}
\usepackage{pst-optexp}
\usepackage{mathtools}
\usepackage{subcaption}
\usepackage[hidelinks]{hyperref}
\usepackage{mdframed}
\usepackage{amsthm}

\usepackage{caption}
\allowdisplaybreaks

\usepackage{graphicx}
\usepackage{dcolumn}
\usepackage{bm}
\usepackage{hyperref}

\usepackage[
margin=0.9in,
]{geometry}

\begin{document}

\baselineskip 16pt

\preprint{APS/123-QED}

\title{Error Free Quantum Target Finding: When Sequential Detection Meets High Dimensional Entanglement} 

\author{Armanpreet Pannu}
\affiliation{The Edward S. Rogers Department of Electrical and Computer Engineering, University of Toronto}

\author{Amr S. Helmy}
\affiliation{The Edward S. Rogers Department of Electrical and Computer Engineering, University of Toronto}

\author{Hesham El Gamal}
\affiliation{School of Electrical and Information Engineering, Faculty of Engineering, The University of Sydney}

\date{\today}

\begin{abstract}
We present a new method for target finding and ranging in Lidar applications using high-dimensional Bell states. Combined with a sequential decision rule, this scheme asymptotically achieves zero error probability with finite energy expenditure. This result stems from the synergy of high-dimensional entanglement suppressing false positives and the sequential decision rule optimizing energy usage. It effectively provides a pathway to unbounded quantum advantage over classical methods and has substantial implications for high-precision sensing in noisy environments.
\end{abstract}

\maketitle

\section{\label{sec:intro}Introduction}
The advancement of quantum technologies has unlocked new possibilities in sensing and detection, enabling performance that surpasses classical limits by exploiting non-classical properties of light~\cite{giovannetti2006quantum,advances_quantum_metrology_2011}. Quantum illumination (QI), or entanglement-enhanced target detection, has emerged as a promising protocol for target detection in noisy environments~\cite{lloyd_original_QI, shapiro_and_lloyd_QI,tan_et_al}. Applications such as LIDAR systems, remote sensing, and ranging rely heavily on precise and efficient detection capabilities, especially when operating in conditions with significant background noise or low target reflectivity~\cite{pirandola2018advances, quantum_lidar_review_Shabir_2024}.

QI has been shown to offer up to a 6 dB advantage in the error exponent~\cite{nair_and_gu} achieved by using a two-mode squeezed vacuum (TMSV) transmitter in the high-noise, low-reflectivity regime \cite{tan_et_al}.  While this represents a notable improvement over classical methods, the full realization in a practical implementation of QI present significant challenges \cite{experimental_QI_italian_2013,experimental_QI_shapiro_2013,experimental_QI_shapiro_2015}. The modest 6 dB gain often does not justify the additional complexity and resource requirements inherent in quantum systems.

Recent efforts have explored using high-dimensional Bell states, which are discrete variable (DV) entangled states, as an alternative to the TMSV QI protocol~\cite{bell_state_QI_pannu}. It was shown that by increasing the number of modes (dimension), the noise could be suppressed asymptotically to zero. This result offers a promising solution for applications where false positives are particularly problematic. However, in a symmetric target detection application, DV states are still unable to surpass the 6 dB quantum advantage limit as increasing the number of modes comes at the cost of reducing the successful detection rate. The overall result in the binary hypothesis QI problem is that DV states are only able to match the performance of QI with TMSV.

In many practical scenarios, the problem extends beyond mere detection to accurately locating a target among multiple possible positions. This is the essence of the target finding problem (and equivalently the ranging problem), which involves identifying the correct location (or equivalently distance) of a known target amidst a multitude of possibilities, each potentially obscured by noise and loss~\cite{target_finding_first_paper}. Classical strategies for target finding and ranging struggle in such noisy environments, often requiring substantial energy expenditure to achieve acceptable error rates, which may be impractical for certain applications in Lidar \cite{lidar_system_architectures_and_circuits,survey_on_lidar_scanning_mechanisms}.

To address these challenges, we propose a novel approach that leverages high-dimensional Bell states combined with a sequential decision-making strategy. By generalizing the DV QI protocol in \cite{bell_state_QI_pannu} to the target finding and ranging problem, we demonstrate that it is possible to asymptotically suppress the error probability to zero with finite energy expenditure, effectively achieving an unbounded quantum advantage. This is accomplished by leveraging the noise suppression capability of high-dimensional DV state transmitters and implementing a sequential detection scheme that minimizes energy use.

Our work represents a significant departure from the perceived 6 dB quantum advantage limit, showing that, for certain applications in sensing, quantum protocols can vastly outperform classical counterparts. This advancement has profound implications for the development of quantum sensing technologies, potentially enabling applications that were previously unattainable due to noise and energy constraints. For instance, in microwave imaging, where the cosmic microwave background introduces substantial noise~\cite{barzanjeh2015microwave}, our protocol could offer unprecedented detection capabilities. The principles demonstrated here may also find uses in problems like secure communication~\cite{shapiro2014secure} and other applications where precision and efficiency are of paramount importance.

\section{Background}
In the context of sensing and LIDAR, target finding and ranging involves probing $D$ positions or distances, one of which contains a target modeled by a noisy bosonic channel, while the remaining positions contain only noise. The noisy bosonic channel is a beamsplitter with reflectivity $\eta$, where the auxiliary port initially contains thermal noise with a mean photon number of $\frac{\mathcal{N}_B}{1-\eta}$. The remaining positions simply replace the incoming state with thermal noise of mean energy $\mathcal{N}_B$, which we will refer to as $\rho_{env}$ from here onwards.

For each of the $D$ possible positions, we probe using a copy of the initial state $\rho_{in}$. Therefore, if the target is at position $x$, the final state after interaction is given by:
\begin{align}
    \rho^{x}= \rho_{pres}^x \otimes \bigotimes_{k\neq x} \rho_{env}^k
\end{align}
Here, $\rho_{pres}$ refers to the signal state $\rho_{in}$ after it passes through the noisy channel and mixes with the noise. The superscripts indicate the different Hilbert spaces corresponding to each possible position.

This problem was first formalized and studied by Zhuang and Pirandola~\cite{target_finding_first_paper} in the general context of channel position finding. In their work, they established the lower bound for the probability of error in a classical target finding (CTF). The optimal probability of error is achieved by probing the possible locations using coherent light and is subject to the lower bound:
\begin{align}
    P^{CTF} \geq \frac{D-1}{2D} \exp \left[-\frac{2\eta  \mathcal{N}_S }{2 \mathcal{N}_B+1}\right] \label{pe_ctf_LB_Pirandola}
\end{align}
Here, $\mathcal{N}_S$ represents the total energy radiated per position which is typically split across $M$ modes. The total energy expended by probing all possible positions is therefore $D \mathcal{N}_S$. The background noise is assumed to be in a thermal state with a mean photon number $\mathcal{N}_B$ per mode.

The same work also compared this classical detection scheme to a quantum-enhanced target-finding scheme. In the quantum-enhanced scheme, each position was probed with the signal from a two-mode squeezed vacuum, while the idler was retained. Under the limit of low energy per mode, $\mathcal{N}_S/M \ll 1$, and a large number of modes per position, $M \gg 1$, the following upper bound was determined for target finding with TMSV transmitter:
\begin{align}
    P^{TMSV} \leq (D-1) \exp\left[-\frac{ \eta \mathcal{N}_S}{1+\mathcal{N}_B}\right] \label{pe_qtf_TMSV_UB_Pirandola}
\end{align}
These two bounds (\ref{pe_ctf_LB_Pirandola}) and (\ref{pe_qtf_TMSV_UB_Pirandola}) suggest that the quantum approach offers no significant advantage over the classical protocol. Interestingly however, it was also shown that a 3dB quantum advantage could be attained using a generalized conditional-nulling receiver in the limit $\mathcal{N}_B \gg 1$.

Although this model and performance bounds apply to both ranging and target detection, they are not optimized for ranging. In this model, each potential target position is probed independently with a separate transmitter however, the most energy-efficient approach to ranging involves probing the target with a single transmitter and determining the time of flight, thereby reducing energy consumption by a factor of $D$. This efficient strategy was explored in \cite{ranging_zhuang}, where it was demonstrated that a quantum advantage of up to 6 dB can be achieved in ranging using a two-mode squeezed vacuum (TMSV) transmitter.

\section{Target Finding with Discrete Variable States}

To address the target finding and ranging problem with discrete variable (DV) states using the model in the previous section, we consider the quantum target detection scheme proposed in \cite{bell_state_QI_pannu}. This work demonstrates that by starting with the $M$-mode Bell state: $\ket{\psi_1}=\sum_{i=1}^M \ket{\mathbf{e}_i,\mathbf{e}_i}$
where $\ket{\mathbf{e}_i,\mathbf{e}_i}$ represents the state with a signal and an idler photon in the $i$-th mode, and selecting an appropriate quantum receiver, one can achieve the same 6 dB error exponent advantage as the well-established protocols using the TMSV. The maximum advantage is attained asymptotically in the limit $M \rightarrow \infty$. More specifically, the probability of a false positive using this protocol can be upper bound by $\frac{1}{M}$, which tends to zero, while the probability of a true positive always remains above $\frac{\eta}{1+\mathcal{N}_B}$. The number of photons transmitted per run is always one regardless of the number of modes.

The ability to suppress false positives to zero becomes increasingly valuable with an adaptive decision rule. It was also shown in \cite{bell_state_QI_pannu} that a simple sequential decision rule can surpass the $6$ dB limit, with the quantum advantage increasing as the prior probability of the target being present rises. This makes DV QI with sequential detection particularly promising for the target finding and ranging problem, where the target is known to exist.

We apply this DV QI protocol to each possible position with the measurements $\{\hat{P}^x, 1-\hat{P}^x\}$ as outlined in \cite{bell_state_QI_pannu}, corresponding to each possible position $x$. The two outcomes represent either the target is present at position $x$ or not present at $x$. Repeating this process $\mathcal{N}_S$ times, we obtain the D-vector $\mathbf{b} = (b_1, \dots, b_D)$, where $b_i$ represents the number of times, out of the $\mathcal{N}_S$ trials, that the measurement at the $i$-th position resulted in a positive outcome. The total energy expenditure of this scheme is $\mathcal{N}_S D$ photons.

In the limit $M \rightarrow \infty$, the probability of a false alarm approaches zero, meaning that if the target is not at position $i$, then $b_i = 0$. Consequently, our decision rule is to simply identify the component of $\mathbf{b}$ that is non-zero and conclude that the target is at that position. An error only occurs if $\mathbf{b} = \mathbf{0}$, indicating that no photons were returned. For each trial, the probability of failed detection, if the target is present, is $\tilde{\eta} = \frac{\eta}{1+\mathcal{N}_B}$, so the probability of error in this case is simply:
\begin{align}
    P_e^{DV,QTF}=\left(1-\tilde{\eta}\right)^{\mathcal{N}_S} \approx \exp \left[ - \frac{\eta \mathcal{N}_S}{1+\mathcal{N}_B}\right]
\end{align}
This result shows no advantage in the error exponent compared to the optimal classical target-finding scheme (\ref{pe_ctf_LB_Pirandola}) and is consistent with the upper bound (\ref{pe_qtf_TMSV_UB_Pirandola}) demonstrated for the TMSV target-finding scheme \cite{target_finding_first_paper}.

To achieve performance beyond classical capabilities, we employ a sequential decision rule. We repeat the measurement until we receive a single positive click, at which point we declare the target to be at that position. The average number of transmissions before a positive click is $\frac{1}{\tilde{\eta}}$, and each transmission uses $D$ photons. Hence, the average energy used is $\frac{D}{\tilde{\eta}}$. By continuing the transmissions until we obtain a click, the probability of missed detection becomes zero. The false alarm rate is suppressed by increasing $M \rightarrow \infty$. The result is a protocol that achieves zero error with finite energy.

In the case where the energy is limited such that $N_{max}$ is the maximum number of transmissions available, the average photon  is given by:
\begin{align}
    \mathcal{N}_S
    &=
    \frac{D}{\tilde{\eta} }\Big(1-(1-\tilde{\eta})^{N_{max}}\Big)
\end{align}
A missed detection occurs only if all $N_{max}$ trails fail and hence the probability of a false alarm becomes
\begin{align}
    P_e^{QTF,Sq} = (1-\tilde{\eta})^{N_{max}} \approx \exp \left[ - \frac{\eta N_{max}}{1+\mathcal{N}_B}\right]
\end{align}
\begin{figure}[h]
    \centering
    \includegraphics[scale=0.5]{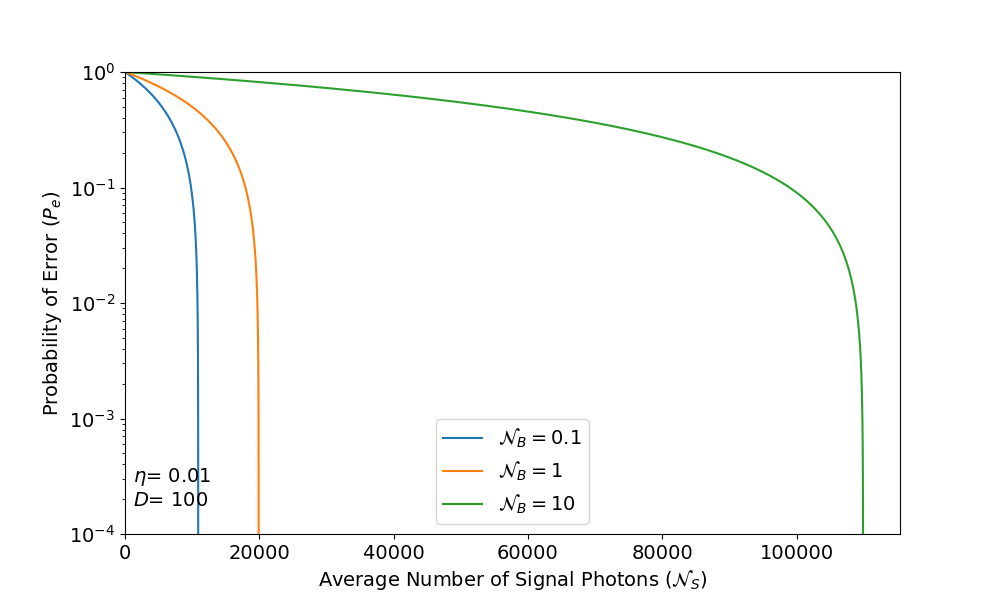}
    \caption{Probability of error for the CPF problem using the $M$-mode maximally entangled discrete variable states in the asymptotic limit $M \rightarrow \infty$.}
    \label{fig:target_finding_infinite_M}
\end{figure}
The overall probability of error is shown in Figure \ref{fig:target_finding_infinite_M}, where we observe that a zero probability of error can be achieved with finite energy.

\section{Target Finding with finite-dimensional Bell states}
While the asymptotic result of $M \rightarrow \infty$ offers valuable insights and highlights the potential of Bell state sensing, it has limited practical relevance, particularly when the claim involves a noise-free sensing protocol. In this section, we consider the case of finite $M$ and examine how the error probability scales.

With finite $M$, the decision rule must be adapted as there is a non-zero probability less than $\frac{1}{M}$ of false detection at each incorrect position. Based on the results from the previous section, we omit the analysis of the non-sequential protocol which fails to provide any advantage over the classical scheme, even in the asymptotic limit. Instead, we focus entirely on a sequential decision rule. 

When there is a possibility of a false alarm at each position, for each transmission, we may observe a click at the correct position, a click at an incorrect position, or clicks at multiple positions. A simple yet robust decision rule is to repeat the measurement until a single position registers $R$ clicks. If $R$ clicks occur for two hypotheses in the same trial, we declare an error. Under this decision rule, the average energy transmitted is:
\begin{align}
\mathcal{N}_S &\leq \sum_{n=R}^\infty n {n-1 \choose R-1} \eta^R (1-\eta)^{n-R}
=
\frac{R}{\eta}   \label{N_S_CTF_R}
\end{align}
The simplest case is $R=1$ where we wait for only a single click and an error occurs only if one of the incorrect positions gets a click before the correct position:
\begin{align}
    P_e^{sq,1C}
    &\leq
    \sum_{n=1}^\infty \left(\frac{(D-1)}{M}\right) (1-\tilde{\eta})^{n-1}
    = \frac{1}{M}\frac{(D-1)}{\tilde{\eta}} 
\end{align}
We achieve an $O\left(\frac{1}{M}\right)$ scaling in the probability of error with a fixed average energy of $\frac{1}{\tilde{\eta}}$. Therefore, by increasing $M$, we can obtain an arbitrarily low probability of error, even if we only wait for a single click.

If we wish to use more energy or are limited in how high we can set $M$ (or both), we need to wait for more clicks (i.e., increase $R$). In this case, the probability of error is the likelihood that any one of the incorrect possible outcomes clicks $R$ times before the correct one does. Since the measurements at each position are independent, we can consider the probability that any individual incorrect position clicks $R$ times before the correct one. Letting $\mathbf{b}(n) = (b_1(n), \dots, b_D(n))$ represent a counter sequence that tracks the number of successes after $n$ trials, and assuming without loss of generality that the correct target position is $x=1$, the probability of a particular incorrect outcome $x \neq 1$ is given by:
\begin{align}
    P_e^{x}= \sum_{n=R}^\infty {n-1 \choose R-1}\left(1-\frac{1}{M}\right)^{n-R}\left(\frac{1}{M}\right)^R Pr[b_1(n)\leq R] \label{eq:pe_DVQTF_finite_M_seq_raw}
\end{align}
This sum is over all $n$ greater than $R$, where $n$ represents the trial number at which the error occurred. For an error to occur on the $n$-th trial, there must be a positive click at position $x$ on that trial, and the previous $(n-1)$ trials must have had $R-1$ clicks. Additionally, the correct position must have accumulated fewer than or equal to $R$ clicks. 

Since each click at the correct position occurs with probability $\tilde{\eta}$, $b_1(n)$ follows a binomial distribution with $n$ trials, and $Pr[b_1(n)\leq R]$ corresponds to the Bernoulli cumulative distribution function (CDF). By bounding the CDF with the Chernoff bound, we can compute the following upper bound for the probability of measuring an incorrect hypothesis (see supplementary materials):
$
    P_e^{x}\leq
    \frac{1}{M^R} C(R)
$
where $C(R) = \frac{e}{R\sqrt{R}\sqrt{2R-1}}\left(\frac{(2R-1)^2}{R^2 \tilde{\eta}}\right)^R$. Note that in this case, the energy expended is represented by the variable $R$, using the relationship given in equation (\ref{N_S_CTF_R}).

With $(D-1)$ incorrect hypotheses, the overall probability of error, assuming $P_e^{x}\ll D-1$ can be bound by:
\begin{align}
   P_e^{sq,RC} \leq\frac{1}{M^R} (D-1)C(R)
\end{align}

\begin{figure}[h]
    \centering
    \includegraphics[scale=0.5]{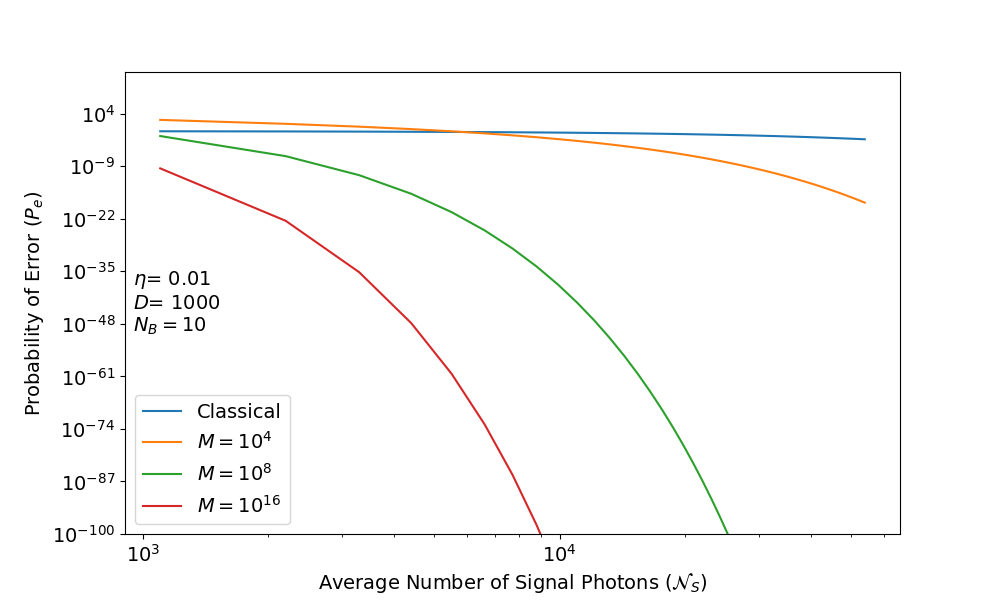}
    \caption{Probability of error for the CPF problem using the $M$-mode maximally entangled discrete variable state. The $x$-axis shows the average photon expenditure calculated using equation (\ref{N_S_CTF_R})} \label{target_finding_finite_M}
\end{figure}

In this case, the overall error probability is $O\left(\frac{D-1}{M^R}\right)$ for a fixed value of $R$. Hence, once again, we can arbitrarily increase $M$ to achieve a zero probability of error but at a faster rate for higher $R$. Alternatively, if $M$ is limited, we can increase $R$ (the energy) to reduce the probability of error, as shown in Figure \ref{target_finding_finite_M}.

\section{Conclusion}
We have shown that high-dimensional quantum entanglement, in conjunction with a sequential decision rule, allows for an unbounded quantum advantage in the target finding and ranging problem in LIDAR applications. This represents a significant advancement over classical schemes and provides a new pathway for quantum enhancement in multi-hypothesis sensing problems where the target is known to exist.

Underlying this result is a sequential decision rule along with the unique properties of the DV QI scheme which leverages high-dimensional entanglement to achieve zero false alarm probability at fixed energy while the successful detection rate stays greater than $\frac{\eta}{1+\mathcal{N}_B}$. This is not possible with a classical scheme where achieving zero false alarm rate requires increasing the energy or sacrificing the detection rate asymptotically to zero.

Our results open exciting possibilities for the development of quantum sensing technologies that can far surpass classical limits. The ability to minimize detection errors while conserving energy has profound implications for applications such as microwave imaging, where the cosmic microwave background presents a problematic amount of noise \cite{barzanjeh2015microwave}, or for general applications requiring high accuracy and precision. Furthermore, the proposed protocol likely admits generalization to non-sensing applications such as secure communication \cite{shapiro2014secure}.

However, implementing this protocol poses several practical challenges. The generation and manipulation of discrete variable high-dimensional entangled states require sophisticated technology, including precise state generation~\cite{advances_high_dim_entaglement_review}, phase-sensitive measurements~\cite{book_quantum_measurement_Braginsky}, and efficient idler storage~\cite{quantum_memories_review}, all of which are still in their infancy. Additionally, implementing the protocol in parallel across multiple positions, as required in the proposed protocol, introduces complexities related to scalability and synchronization.

The unbounded quantum advantage demonstrated in this manuscript should motivate further research into addressing the practical challenges with implementation, and theoretical research into new applications or easier-to-implement protocols based on the principles of high-dimensional entanglement and sequential detection.

\bibliography{bibliography}

\end{document}